\documentclass[aps,pra,reprint,groupedaddress,showpacs]{revtex4-1}
\usepackage{bm}
\usepackage[dvipdfmx]{graphicx}
\usepackage{times}
\usepackage{color}
\usepackage{mathtools}
\usepackage{setspace}
\usepackage{multirow}
\usepackage{here}

\begin{document}

\title{Requirements for fault-tolerant quantum computation with cavity-QED-based atom-atom gates mediated by a photon with a finite pulse length}

\author{Rui \surname{Asaoka}$^{1}$\email[]{rui.asaoka.st@hco.ntt.co.jp}, 
Yuuki \surname{Tokunaga}$^{1}$,
Rina \surname{Kanamoto$^{2}$},
Hayato \surname{Goto}$^{3}$, and 
Takao \surname{Aoki}$^{4}$}

\affiliation{$^{1}$Computer and Data Science Laboratories, NTT Corporation, Musashino 180-8585, Japan\\
$^{2}$Department of Physics, Meiji University, Kawasaki, Kanagawa 214-8571, Japan\\
$^{3}$Corporate Research and Development Center, Toshiba Corporation, Kawasaki, Kanagawa 212-8582, Japan\\
$^{4}$Department of Applied Physics, Waseda University, Shinjuku, Tokyo 169-8555, Japan\\
}

\date{\today}


\begin{abstract}
We analyze the requirements for fault-tolerant quantum computation with atom-atom gates based on cavity quantum electrodynamics (cQED) mediated by a photon with a finite pulse length. For short photon pulses, the distorted shape of the reflected pulses from the cQED system is a serious error source.
In the previous study by Goto and Ichimura [Phys. Rev. A {\bf 82}, 032311 (2010)], only the photon loss is minimized without considering the shape distortion to optimize the system parameters. Here we show an improved optimization method to minimize the infidelity due to the shape distortion and the photon losses in a well-balanced manner for the fault-tolerant scheme using probabilistic gates [Phys. Rev. A {\bf 80}, 040303(R) (2009)]. Under this optimization, we discuss the FTQC requirement for short pulses.
Finally, we show that reducing the cavity length is an effective way to reduce the errors of this type of gate in the case of short photon pulses.
\end{abstract}

\maketitle

\section{Introduction} 
Cavity quantum electrodynamics (cQED) is an important platform for quantum computing because it enables atoms and light to couple at the single-photon level. So far, many schemes based on cQED systems, such as single-photon sources \cite{kuhn2010,law,kuhn1999,maurer,vasilev}, non-demolition measurements \cite{geremia,reiserer2013,hosseini}, and two-qubit gates \cite{zheng,duan2004,duan2005,xiao2004,xiao2004_2,koshino,reiserer2014,hacker,daiss}, have been proposed and demonstrated as a means of implementing quantum computers \cite{reiserer2015}. The ultimate goal of these theoretical and experimental efforts is the realization of fault-tolerant quantum computing (FTQC) \cite{nielsen2000}.

It is well known that FTQC can be achieved if the error probability per operation used for scalable quantum computation is lower than a certain threshold \cite{nielsen2000,shor}. In particular, sufficiently small error probability of two-qubit gates is often the most stringent requirement for FTQC.

Previously, the requirement for the FTQC using probabilistic gates based on cQED \cite{goto2009} has been investigated in the long pulse limit \cite{goto2010}. In this case, cQED systems can be optimized by simply minimizing the probability of photon loss through dissipative channels.
However, the FTQC requirement may become more severe for short pulses, because of the error induced by distortion of the shape of the pulses through reflections from cQED systems.
On the other hand, pulses should be shortened for faster gate operations and shorter delay lines.

In this paper, we investigate the FTQC requirements for cQED-based quantum gates with finite pulse lengths and optimize the cavity parameters, mainly focusing on the external coupling rate. The optimization is performed so that the FTQC requirements for the FTQC scheme proposed in Ref.\,\cite{goto2009} are most likely to be met. That is, the photon loss probability and the pulse-distortion error probability are minimized in the best-balanced manner for the FTQC scheme. Our optimization greatly relaxes the FTQC requirements in some parameter regions, compared with the previous optimization method that simply minimizes the photon loss \cite{goto2010}.

In the long pulse limit, the error probability of the controlled phase flip (CPF) gate can be represented by a single dimensionless value called {\it internal cooperativity} \cite{goto2010}. In contrast, we find that the performance of the CPF gate is not characterized only by the internal cooperativity when the photon pulse length is shorter than ${\sim 1/\kappa_{\rm int}}$, where $\kappa_{\rm int}$ is the cavity field decay rate due to undesirable scattering and absorption inside the cavity. Finally, we show that reducing the cavity length is effective at meeting the FTQC requirements when the photon pulse length is shorter than $1/\kappa_{\rm int}$.

This paper is organized as follows. In Sec. II, we briefly explain the CPF gate scheme based on cQED as proposed in Ref.\,\cite{duan2005}. In Sec. III, we classify the errors in the CPF gate. In Secs. IV and V, we discuss the requirements for FTQC with the cQED scheme in the long pulse limit and for a finite pulse length, respectively. Our conclusions are presented in Sec. VI.
%
\section{Cavity-QED based controlled phase flip gate} 
\label{gatesec}
\begin{figure}
\includegraphics[clip,width=8cm]{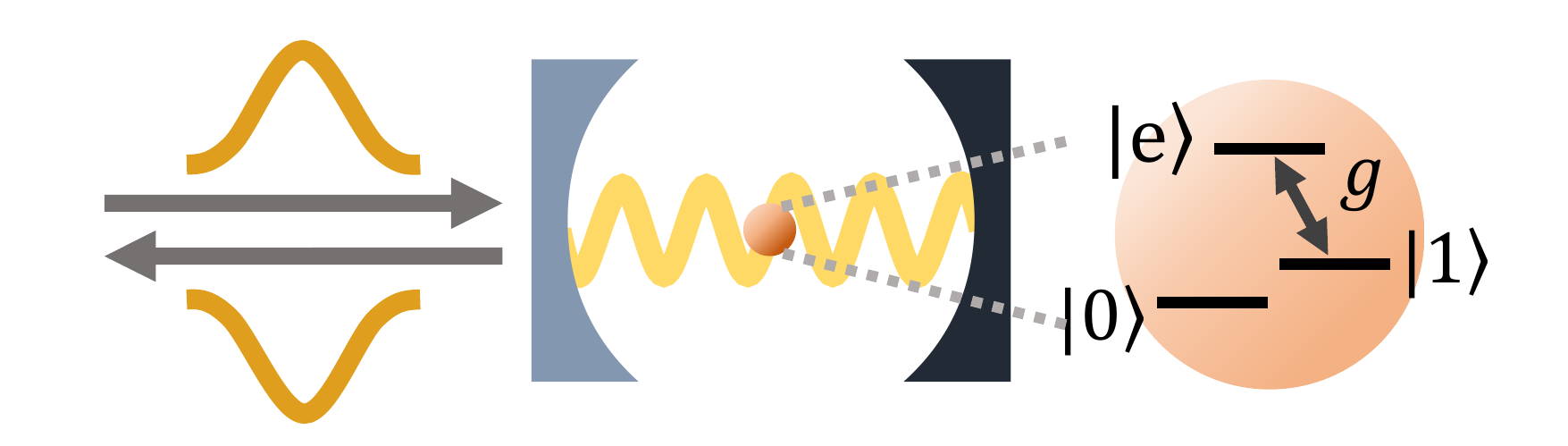}
\caption{Schematic of the interaction between a three-level atom in a single-sided cavity and a single-photon pulse.}
\label{int}
\end{figure}

The CPF gate between atom qubits proposed in Refs.\,\cite{duan2005,xiao2004,xiao2004_2} utilizes {\it selective $\pi$-phase flip reflection} in the interaction between a three-level atom in a single-sided cavity and a single-photon pulse. First, we explain the phase flip mechanism \cite{duan2004}. A qubit is represented by the ground states of an atom, $|0\rangle_{\rm a}$ and $|1\rangle_{\rm a}$, and the transition between $|1\rangle_{\rm a}$ and an atomic excited state $|{\rm e}\rangle_{\rm a}$ is resonant with a cavity (Fig.\,\ref{int}). The Hamiltonian of a cQED system with input and output photon pulses is given by
%
\begin{align}
\nonumber
H &= \hbar\omega_{\rm c}c^{\dagger}c
+ \hbar\omega_{\rm 1e}|{\rm e}\rangle_{\rm aa}\langle{\rm e}| - \hbar\omega_{01}|{0}\rangle_{\rm aa}\langle{0}| \\
\nonumber
&+ \hbar\int_{-\infty}^{\infty}d\omega\omega a^{\dagger}(\omega)a(\omega)  \\
\nonumber
&+ i\hbar g\left(c^{\dagger}|1\rangle_{\rm aa}\langle{\rm e}| - c|{\rm e}\rangle_{\rm aa}\langle1|\right) \\
&+ i\hbar\sqrt{\frac{\kappa_{\rm ext}}{\pi}}\int_{-\infty}^{\infty}d\omega\left[c^{\dagger}a(\omega) - a^{\dagger}(\omega)c\right]   ,
\end{align}
where $c (c^{\dagger})$ and $a (a^{\dagger})$ are the annihilation (creation) operators of a photon in the cavity and of a pulse photon, respectively.
The entire state of the cQED system and a photon pulse is expressed as
%
\begin{align}
\nonumber
|\Psi(t)\rangle &= |0\rangle_{\rm a}|0\rangle_{\rm c}\int^{\infty}_{-\infty}d\omega f_0(\omega,t)a^{\dagger}(\omega)|0\rangle_{\rm p} \\
\nonumber
                    &+ c_0(t)|0\rangle_{\rm a}|1\rangle_{\rm c}|0\rangle_{\rm p}  \\
\nonumber
                    &+ |1\rangle_{\rm a}|0\rangle_{\rm c}\int^{\infty}_{-\infty}d\omega f_1(\omega,t)a^{\dagger}(\omega)|0\rangle_{\rm p} \\
\nonumber
                    &+ c_1(t)|1\rangle_{\rm a}|1\rangle_{\rm c}|0\rangle_{\rm p}  \\
                    &+ d(t)|{\rm e}\rangle_{\rm a}|0\rangle_{\rm c}|0\rangle_{\rm p},
\end{align}
where $f_{0(1)}(\omega, t)$ is the probability amplitude of the incident single-photon pulse when the atom is in the $|0\rangle_{\rm a} (|1\rangle_{\rm a})$ state. The subscripts `c' and `p' denote `cavity' and `pulse', respectively. The Fock space of the cavity field is truncated at one, assuming that the pulse includes only a single photon.
We also take into account the following dissipative processes: atomic spontaneous emission with a (polarization) decay rate $\gamma$, cavity field decay with a decay rate $\kappa_{\rm ext}$ associated with the extraction of a cavity photon to the desired external mode via transmission of the mirror, and other undesirable cavity field decay due to the imperfection of the cavity with the rate $\kappa_{\rm int}$. The total cavity decay rate is given by ${\kappa=\kappa_{\rm ext}+\kappa_{\rm int}}$.

Finally, we derive the equations of motion for the cQED system with dissipative channels and the relations between the input pulse and output pulse \cite{walls}:
%
\begin{align}
\label{em1}
&\dot{C_0} =  - \kappa C_0 - \sqrt{\frac{\kappa_{\rm ext}}{\pi}}\int_{-\infty}^{\infty}d\omega F_0^{\rm in}(x=0)e^{-i\omega t}   , \\
\label{em2}
&\dot{C_1} = gD - \kappa C_1 - \sqrt{\frac{\kappa_{\rm ext}}{\pi}}\int_{-\infty}^{\infty}d\omega F_1^{\rm in}(x=0)e^{-i\omega t}   , \\
\label{em3}
&\dot{D} = - gC_1 - \gamma D   , \\
\label{em4}
&F_0^{\rm out}(x=0) = F_0^{\rm in}(x=0) + \sqrt{\frac{\kappa_{\rm ext}}{\pi}}\int_{0}^{t}dt' C_0e^{-i\omega t'}   , \\
\label{em5}
&F_1^{\rm out}(x=0) = F_1^{\rm in}(x=0) + \sqrt{\frac{\kappa_{\rm ext}}{\pi}}\int_{0}^{t}dt' C_1e^{-i\omega t'}   ,
\end{align}
where we have introduced the probability amplitudes in the rotating frame: ${C_0 = e^{i(\omega_{\rm c}-\omega_{\rm 01})t}c_0}$, ${C_1 = e^{i\omega_{\rm c}t}c_1}$, ${D = e^{i\omega_{\rm 1e}t}d}$, ${F_0^{\rm in} = f_0(\omega, t=0)e^{i(\omega_{\rm c} - \omega_{01})t}}$, ${F_1^{\rm in} = f_1(\omega,t=0)e^{i\omega_{\rm c} t}}$, ${F_0^{\rm out} = f_0(\omega,t)e^{i(\omega_{\rm c} - \omega_{01})t}}$, and ${F_1^{\rm out} = f_1(\omega,t)e^{i\omega_{\rm c} t}}$. We set ${x=0}$ as the position of the input mirror; notation ${x=0}$ is omitted in the following discussions. Equations (\ref{em4}) and (\ref{em5}) represent the interference between the input pulse and the cavity field, which induces interesting phenomena such as phase flip and resonant tunneling. The norm of the entire state $\langle\Psi(t)|\Psi(t)\rangle$ decreases over time via the dissipative terms, and the decrease in the norm corresponds to the photon loss probability.

One can understand the basic mechanism of the CPF gate by analyzing these equations \cite{duan2004}. Assuming that the photon pulse is sufficiently long, we can make the approximations $\dot{C_0} \simeq 0$, $\dot{C_1} \simeq 0$, and $\dot{D} \simeq 0$. Accordingly, we obtain simple relations between the input and output pulses:
%
\begin{align}
\label{f0}
&F_0^{\rm out} = L_0F_0^{\rm in}   , \\
\label{f1}
&F_1^{\rm out} = L_1F_1^{\rm in}   ,
\end{align}
where
\begin{align}
\label{l0}
&L_0 = \frac{-\kappa_{\rm ext} + \kappa_{\rm int} - i\Delta}{\kappa_{\rm ext} + \kappa_{\rm int} - i\Delta}   , \\
\label{l1}
&L_1 = \frac{-\kappa_{\rm ext} + \kappa_{\rm int} -i\Delta + \frac{g^2}{\gamma-i\Delta}}{\kappa_{\rm ext} + \kappa_{\rm int} -i\Delta + \frac{g^2}{\gamma-i\Delta}}   .
\end{align}
Here, ${\Delta = \omega - \omega_{\rm c}}$ is the detuning. To realize the CPF gate, only the phase of the atomic state $|0\rangle_{\rm a}$ needs to change by $\pi$, namely ${L_0 = -1}$ and ${L_1 = 1}$. We call this selective $\pi$-phase reflection.  
For example, in the long pulse limit, where the input photon includes a single mode ${\omega = \omega_{\rm c}}$, selective $\pi$-phase reflection is achieved when the condition
\begin{align}
{g^2/\gamma \gg \kappa_{\rm ext} \gg \kappa_{\rm int}}
\label{hf}
\end{align}
is satisfied.
%
%
%

\begin{figure}
\includegraphics[clip,width=8cm]{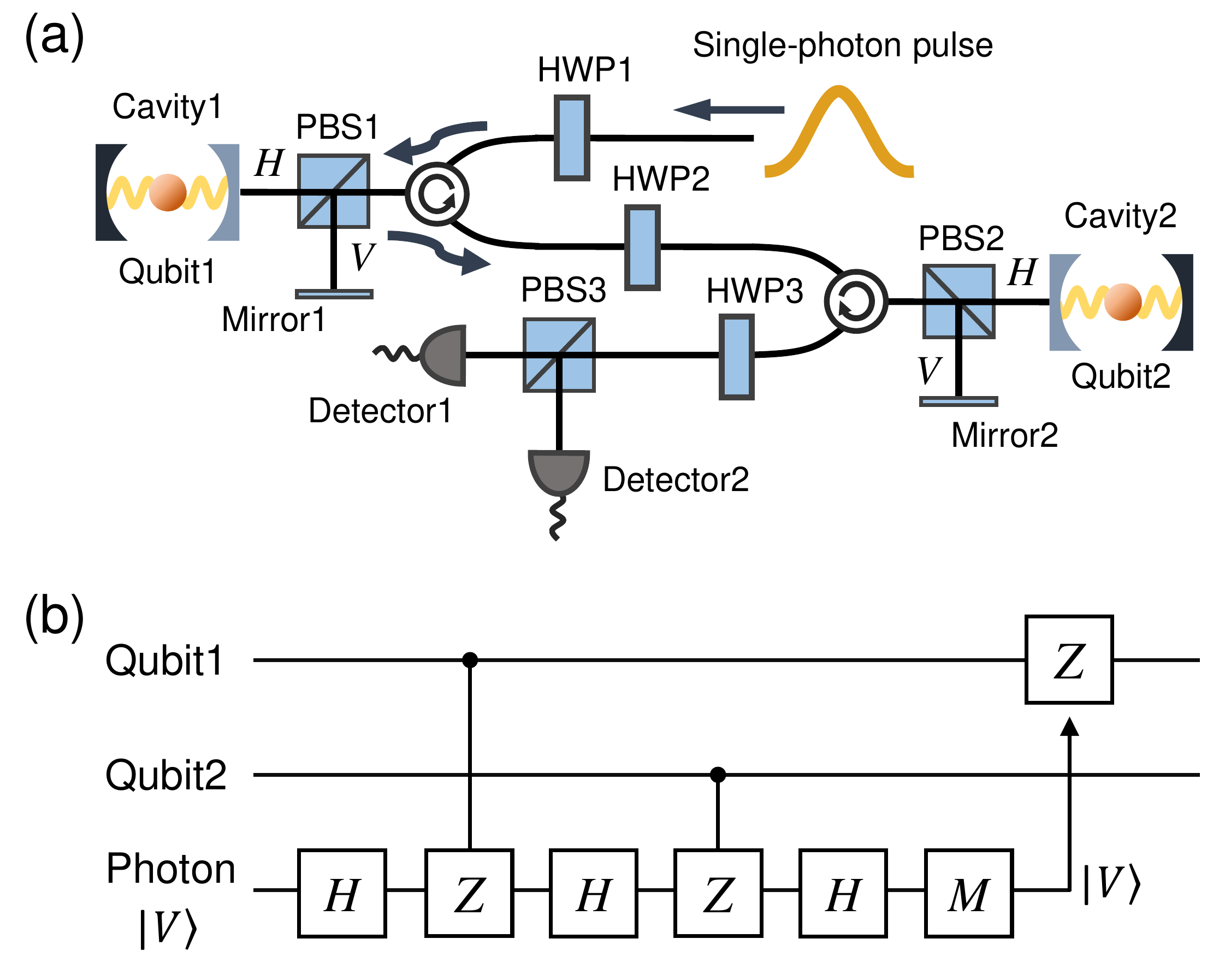}
\caption{(a) Schematic diagram of the CPF gate between atom qubits. (b) A quantum circuit representing the CPF gate in (a). $Z$ and $H$ denote Pauli-$Z$ and Hadamard gates, respectively. $Z$ with a vertical line denotes the CPF gate between an atom qubit and a photon qubit. The rightmost $Z$ at the end of the arrow means that a Pauli-$Z$ gate is performed only when the measurement result is $|V\rangle_{\rm p}$.}
\label{gate}
\end{figure}

The CPF gate between atom qubits is formed by combining the selective $\pi$-phase flip reflection technique with linear optical devices and a feedback operation. The setup of the CPF gate between atom qubits proposed in Ref.\,\cite{duan2005} is illustrated in Fig.\,\ref{gate}(a). A single-photon pulse carries a qubit represented by its polarization states $|V\rangle_{\rm p}$ and $|H\rangle_{\rm p}$. Initially the single-photon pulse is in the $|V\rangle_{\rm p}$ state, and is incident on the cavities only when its polarization is $|H\rangle_{\rm p}$. The polarization of the photon pulse is measured by the polarized beam splitter (PBS3) and the photodetectors, and a phase flip gate $\sigma^z$ is performed on the atom qubit 1 if detector 2 clicks; that is, the photon state is in the $|V\rangle_{\rm p}$ state. Figure \ref{gate}(b) shows the quantum circuit representing the whole operation in Fig.\,\ref{gate}(a). The Hadamard gates on the photonic qubit are performed with the half-wave plates (HWP 1,2,3). This circuit would be equivalent to the CPF gate between qubits 1 and 2 and operates ideally in the long pulse limit if there were no dissipation processes. However, the dissipation processes and the finite pulse length effect effectively prevent ideal gate operation. The key point of this gate is that the photon loss during the CPF gate operation can be naturally treated as a failure event of the probabilistic two-qubit gates, which can be detected by the photon detectors. Hence FTQC schemes for probabilistic two-qubit gates can be applied to quantum computing with this gate.
%
%
%
\section{Fault-tolerant quantum computing scheme and errors in the controlled phase flip gate}
\label{errorsec}
In Ref.\,\cite{goto2010}, Goto and Ichimura applied the FTQC scheme for probabilistic two-qubit gates \cite{goto2009} to quantum computing based on the atom-atom gate explained in the previous section. As a result, the requirements for FTQC based on cQED were greatly relaxed compared with the case that the standard FTQC schemes based on concatenated code \cite{knill} and surface code \cite{raussendorf} are simply applied.

In the FTQC scheme in Ref.\,\cite{goto2010}, the errors in the CPF gate operation are categorized into two types: {\it photon loss} and {\it conditional error} \cite{expla}. Here, photon loss means that no photon is detected by either detector. Thus, photon loss events can be eliminated by postselecting the events that auxiliary photons are detected. Conditional error is defined as unheralded error, which remains when a photon is detected by one of the photon detectors. This error accumulates during successful postselections. The fault-tolerant threshold of the conditional error, therefore, depends on the photon loss probability \cite{goto2009}. The photon loss probability $p_{\rm l}$ and the conditional error probability $p_{\rm c}$ are formulated as ${p_{\rm l}\equiv 1 - \langle\Psi(t)|\Psi(t)\rangle / \langle\Psi(0)|\Psi(0)\rangle}$ and ${p_{\rm c}\equiv 1 - F = 1 - |\langle\Psi_{\rm id}(t)|\Psi(t)\rangle|^2 / \langle\Psi(t)|\Psi(t)\rangle}$, where $\Psi_{\rm id}(t)$ represents the ideal state function at time $t$, and $F$ is the fidelity renormalized at time $t$. The renormalization eliminates the amount of the photon loss measured in terms of the infidelity.
%
\section{Long pulse limit}
\label{lonlim}
%
Here, we discuss the requirements for FTQC in the case of an optimized cQED system in the long pulse limit, where the detuning $\Delta$ in Eqs. (\ref{f0})-(\ref{l1}) is zero. The long pulse limit has been already discussed in Ref.\,\cite{goto2010}, but we will review here it in the case of a broader range of system parameters, in order to compare it with the case of the finite pulse length in Sec.\,\ref{fin}.
%
\subsection{Optimization by the tuning external coupling rate}
\label{opt1}
In the long pulse limit, conditional errors are caused by the unbalanced photon loss between polarization states and between $|0\rangle_{\rm a}|H\rangle_{\rm p}$ and $|1\rangle_{\rm a}|H\rangle_{\rm p}$ states. In the CPF gate scheme, only the $|H\rangle_{\rm p}$ state is incident on the cavity, which leads to unbalanced photon loss between the different polarization states. Additionally, there is a $g^2/\gamma$ difference between the reflection coefficients of $|0\rangle_{\rm a}|H\rangle_{\rm p}$ and $|1\rangle_{\rm a}|H\rangle_{\rm p}$, as can be seen in Eqs.\,(\ref{l0}) and (\ref{l1}). 
The effect of these unbalanced photon losses remains even after the postselection and reduces the fidelity in the form of a conditional errors.

These errors occurring during the CPF gate operation can be minimized by properly adjusting the system parameters. In cQED systems, the external coupling rate $\kappa_{\rm ext}$ is proportional to the transmittance of the mirror, which is relatively easy to adjust. In the long pulse limit, an appropriate choice of $\kappa_{\rm ext}$ can almost surely minimize both the error probabilities of the photon loss and the conditional error.

\begin{figure}
\includegraphics[clip,width=8cm]{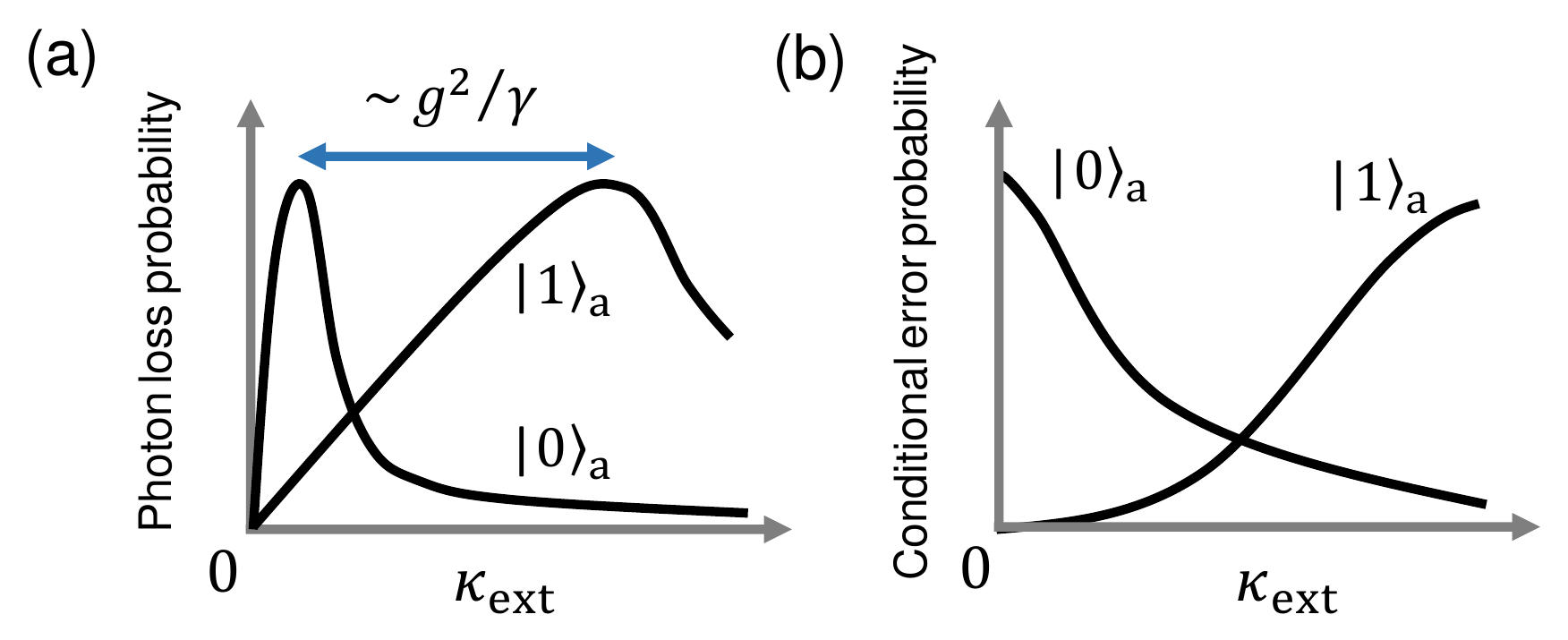}
\caption{Typical dependencies of (a) photon loss probability and (b) conditional error probability on the external coupling rate $\kappa_{\rm ext}$ in the long pulse limit, when the atom is in the $|0\rangle_{\rm a}$ state or $|1\rangle_{\rm a}$ state.}
\label{err2}
\end{figure}

Typical dependencies of the photon loss probability and the conditional error probability on $\kappa_{\rm ext}$ are illustrated in Fig.\,\ref{err2}. These dependencies are straightforwardly explained by Eqs.\,(\ref{f0})-(\ref{l1}).
The photon loss probability in the CPF gate operation between an atom and a photon is calculated by
%
\begin{align}
\label{pl1}
p_{\rm l}^{(0)} = 1-|L_0|^2   ,\\
\label{pl2}
p_{\rm l}^{(1)} = 1-|L_1|^2   ,
\end{align}
where $p_{\rm l}^{(0)}$ and $p_{\rm l}^{(1)}$ denote the photon loss probabilities when an atom is in the $|0\rangle_{\rm a}$ state or $|1\rangle_{\rm a}$ state. Since $L_0$ and $L_1$ depend on $\kappa_{\rm ext}$ as in Eqs.\,(\ref{l0}) and (\ref{l1}), $p_{\rm l}^{(0)}$ and $p_{\rm l}^{(1)}$ have peaks at ${\kappa_{\rm ext}=\kappa_{\rm int}}$ and ${\kappa_{\rm ext}=\kappa_{\rm int}+g^2/\gamma}$, respectively, at which the output pulse completely disappears due to interference between the input pulse and the cavity field. Hence, the total loss probability reaches a minimum between the peaks. The photon loss probability is small when $\kappa_{\rm ext}$ is quite small or large, but we will exclude those regions because the conditional probability is quite high, as shown below.

As illustrated in Fig.\,\ref{err2}(b), the conditional error shows a monotonous behavior of the probability of the unbalanced-photon-loss error. The conditional errors in the CPF gate operation between an atom and a photon are calculated as follows:
%
\begin{align}
\label{pc1}
p_{\rm c}^{(0)} = 1 - \frac{1}{2}\frac{(1-L_{0})^2}{1+L_{0}^2}   ,\\
\label{pc2}
p_{\rm c}^{(1)} = 1 - \frac{1}{2}\frac{(1+L_{1})^2}{1+L_{1}^2}   ,
\end{align}
where $p_{\rm c}^{(0)}$ and $p_{\rm c}^{(1)}$ denote the conditional errors when an atom is in the $|0\rangle_{\rm a}$ state or $|1\rangle_{\rm a}$ state. These errors are calculated using the initial state $\frac{1}{\sqrt{2}}\left(|H\rangle_{\rm p} + |V\rangle_{\rm p}\right)|0,1\rangle_{\rm a}$ and the output state $\frac{1}{\sqrt{2}}\left(L_{0,1}|H\rangle_{\rm p} + |V\rangle_{\rm p}\right)|0,1\rangle_{\rm a}$. As $\kappa_{\rm ext}$ increases, the conditional error probability $p_{\rm c}^{(0)}$ monotonically decreases, whereas $p_{\rm c}^{(1)}$ monotonically increases. Hence, the total conditional error probability reaches a minimum around the intersection of these error curves.
%
%

Goto and Ichimura derived a formula for the external coupling rate that minimizes the photon loss \cite{goto2010}:
\begin{align}
\kappa_{\rm ext}^{\rm loss} = \kappa_{\rm int}\sqrt{1+2C_{\rm int}}  .
\label{kex_goto}
\end{align}
Here, {\it internal cooperativity} $C_{\rm int}$ is defined in terms of the system parameters, ${C_{\rm int} \equiv g^2/(2\kappa_{\rm int}\gamma)}$.
The formula (\ref{kex_goto}) was derived from the optimal condition ${L_0 = -L_1}$ to minimize the following photon loss probability:
\begin{align}
\nonumber
p_{\rm l} = 1- \frac{1}{4}\left\{|\alpha_{0,0}|^2\left[L_0^2(L_0-1)^2 + (L_0+1)^2\right] \right. \\
\nonumber
\left. + |\alpha_{0,1}|^2\left[L_1^2(L_0-1)^2 + (L_0+1)^2\right] \right. \\
\nonumber
\left. + |\alpha_{1,0}|^2\left[L_0^2(L_1-1)^2 + (L_1+1)^2\right] \right. \\
\left. + |\alpha_{1,1}|^2\left[L_1^2(L_1-1)^2 + (L_1+1)^2\right] \right\} ,\hspace{-3.2mm}
\label{pl}
\end{align}
where $\alpha_{i,j}$ are the coefficients of the initial state of the two-qubit system ${|\psi_0\rangle = \sum_{i,j=0,1}\alpha_{i,j}|i\rangle_1|j\rangle_2}$.

Furthermore, in Ref.\,\cite{goto2010}, they numerically showed that the conditional error also takes a minimal value around ${\kappa_{\rm ext} = \kappa_{\rm ext}^{\rm loss}}$. This can be intuitively understood from the fact that the unbalanced-photon-loss error, which is originally caused by photon loss, is the major error type within the conditional error. Accordingly, any type of error in the long pulse limit takes a minimal value around ${\kappa_{\rm ext} = \kappa_{\rm ext}^{\rm loss}}$. This means that $\kappa_{\rm ext}^{\rm loss}$ is optimal not only for the photon loss but also for FTQC. Thus, we can choose $\kappa_{\rm ext}^{\rm loss}$ as an approximate optimal value of $\kappa_{\rm ext}$. As shown in the following subsection, when the external coupling rate is $\kappa_{\rm ext}^{\rm loss}$, the FTQC requirements can be simply represented by the internal cooperativity.
%
\subsection{FTQC requirements}
\label{req1}
\begin{figure}
\includegraphics[clip,width=8cm]{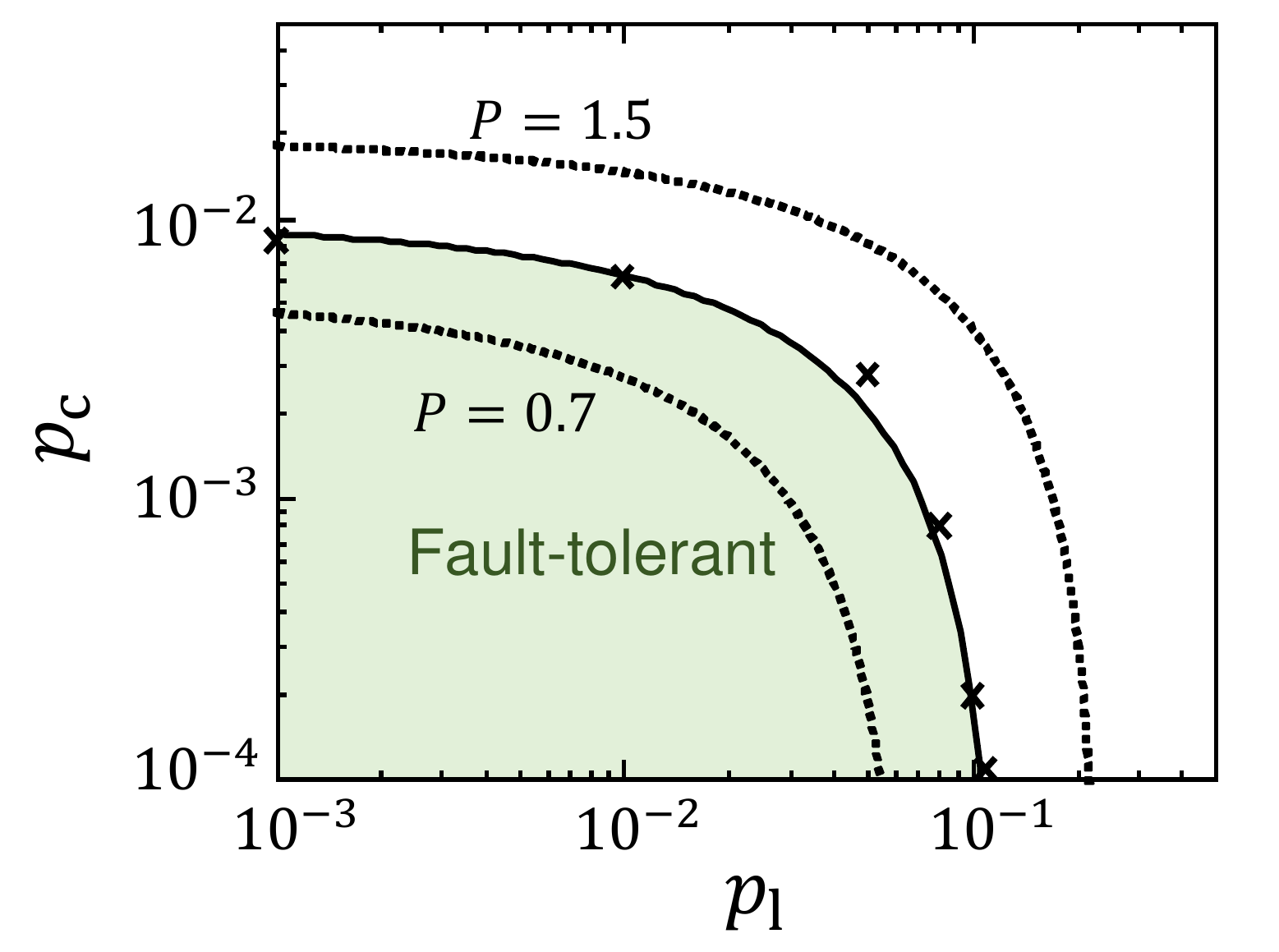}
\caption{Threshold of the FTQC scheme of Ref.\,\cite{goto2009} (crosses). The solid curve represents the FTQC fitted function ${\frac{60}{17}p_{\rm l}^{0.59} + \frac{260}{17}p_{\rm c}^{0.59} = 1}$. The dotted curves denote the average errors for the two values of the FTQC error parameter ${P=0.7}$ and $1.5$.}
\label{thr}
\end{figure}

In Ref.\,\cite{goto2009}, the threshold values of the conditional error probability were derived for several photon loss probabilities (see Fig.\,\ref{thr}). To determine the FTQC requirements for the cQED parameters under the optimization, we can plot a threshold curve over the $p_{\rm l}$-$p_{\rm c}$ map by interpolating the threshold values. We assume that the fitting function for the threshold curve is of the form ${ap_{\rm l}^d + bp_{\rm c}^d = 1}$; the fitted parameters are ${a=60/17}$, ${b=260/17}$, and ${d=0.59}$.
Using this result, we introduce a convenient FTQC error parameter:
\begin{align}
P\equiv \frac{60}{17}\bar{p_{\rm l}}^{0.59} + \frac{260}{17}\bar{p_{\rm c}}^{0.59},
\label{ftqcp}
\end{align}
where $\bar{p_{\rm l}}$ and $\bar{p_{\rm c}}$ are the average photon loss probability and the average conditional error probability, respectively, over the various initial states. As shown in Fig.\,\ref{thr}, if ${P \leq 1}$, a quantum computation with the average errors is fault-tolerant.
The average probabilities are calculated from the following formulas, ${\bar{p_{\rm l}} = 1 - \prod_{i,j}\int d\alpha_{i,j}\langle\Psi(t)|\Psi(t)\rangle/\prod_{i,j}\int d\alpha_{i,j}\langle\Psi(0)|\Psi(0)\rangle}$ and ${\bar{p_{\rm c}} = 1-(4\bar{F}_{\rm max}+1)/5}$, where $\bar{F}_{\rm max}$ denotes the average gate fidelity for the maximally entangled states \cite{nielsen2002}.
We use the formula of fidelity in the long pulse limit to calculate $\bar{F}_{\rm max}$:
\begin{widetext}
\begin{align}
\label{pc}
p_{\rm c} &= 1 - \frac{\left||\alpha_{0,0}|^2L_0(L_0-1) - |\alpha_{0,1}|^2L_1(L_0-1) + |\alpha_{1,0}|^2(L_1+1) + |\alpha_{1,1}|^2(L_1+1)\right|^2}
{|\alpha_{0,0}|^2\eta_1 + |\alpha_{0,1}|^2\eta_2 + |\alpha_{1,0}|^2 \eta_3 + |\alpha_{1,1}|^2 \eta_4} , \\
\eta_1 &= L_0^2(L_0-1)^2 + (L_0+1)^2 , \\
\eta_2 &= L_1^2(L_0-1)^2 + (L_0+1)^2 , \\
\eta_3 &= L_0^2(L_1-1)^2 + (L_1+1)^2 , \\
\eta_4 &= L_1^2(L_1-1)^2 + (L_1+1)^2 .
\end{align}
\end{widetext}

\begin{figure}
\includegraphics[clip,width=8cm]{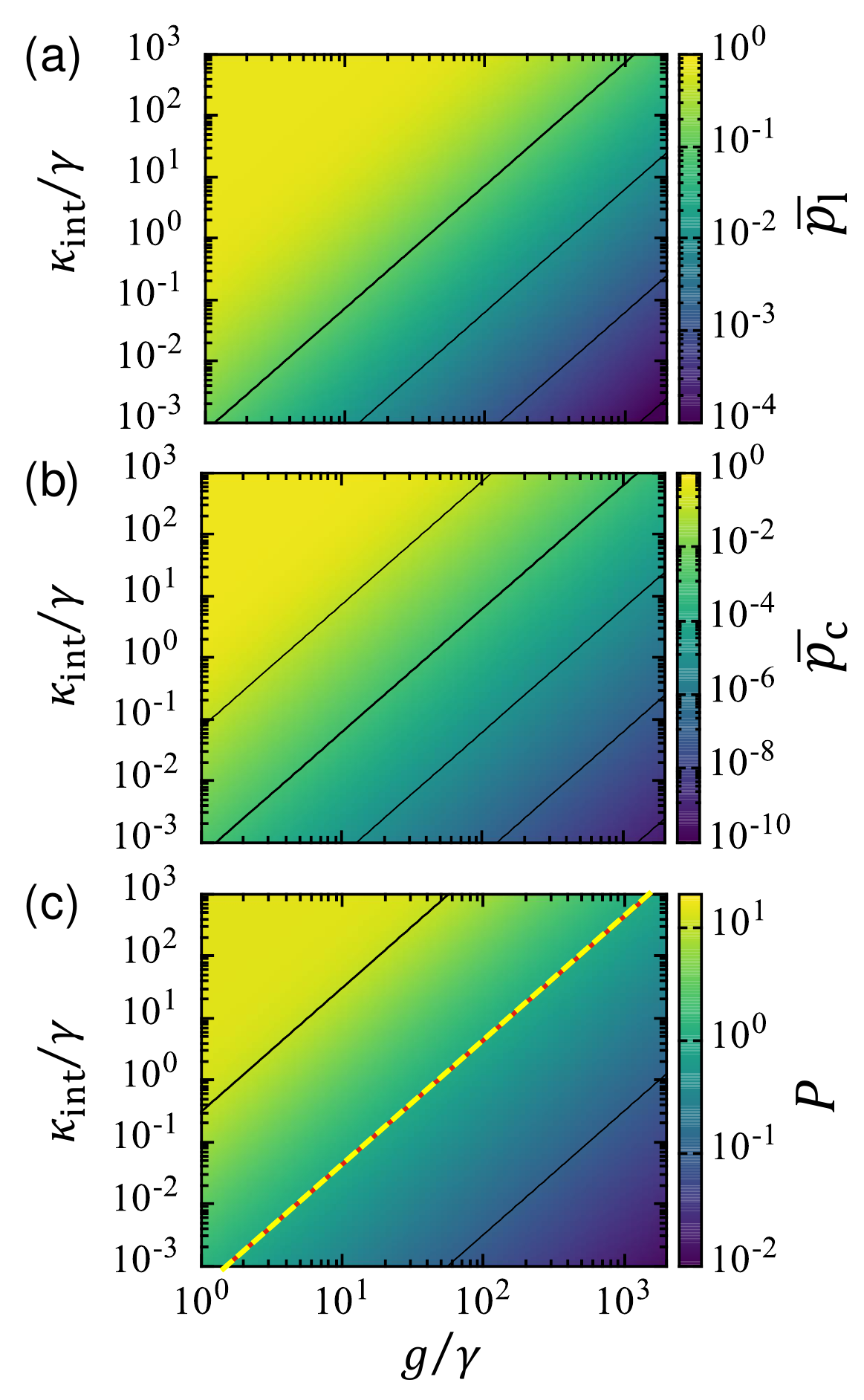}
\caption{(a) Average loss probability $\bar{p_{\rm l}}$, (b) average conditional error probability $\bar{p_{\rm c}}$, and (c) FTQC error parameter $P$ with $\kappa_{\rm ext} = \kappa_{\rm ext}^{\rm loss}$ in the $(g/\gamma, \kappa_{\rm int}/\gamma)$ plane. Solid lines are contours. The dashed line in (c) is the line on which $C_{\rm int}$ is $1130$; it falls along the ${P=1}$ line, namely the FTQC threshold line.}
\label{gkin}
\end{figure}

Fig.\,\ref{gkin} shows the average error probabilities and FTQC error parameter $P$ in Eq.\,(\ref{ftqcp}) in the $(g/\gamma, \kappa_{\rm int}/\gamma)$ plane, with $\kappa_{\rm ext}$ set to $\kappa_{\rm ext}^{\rm loss}$.
Note that the contours of $\bar{p_{\rm l}}$, $\bar{p_{\rm c}}$, and $P$ are lines with gradient 2 on a log-log plot. This means that $\bar{p_{\rm l}}$ and $\bar{p_{\rm c}}$ are expressed only with $C_{\rm int}$ (then $P$ is naturally expressed only with $C_{\rm int}$), because the line on which $C_{\rm int}$ is constant is represented as ${\kappa_{\rm int}/\gamma = \alpha(g/\gamma)^2}$ ($\alpha$ is constant).
In fact, Ref.\,\cite{goto2010} showed that the photon loss probability is proportional to $1/\sqrt{C_{\rm int}}$ for sufficiently large $C_{\rm int}$, and the conditional error probability is also proportional to the inverse of the internal cooperativity when ${\kappa_{\rm int}/\gamma = 1}$. Thus, the value of $C_{\rm int}$ determines the performance of the CPF gate. The reason why the system is not characterized by the general cooperativity ${C \equiv g^2/(2\kappa\gamma)}$ is that simply reducing $\kappa_{\rm ext}$ is not always the best way for atom-photon quantum gates.
%
%
%

We illustrate the line on which $C_{\rm int}$ is $1130$ in Fig.\,\ref{gkin}(c); it falls along the ${P=1}$ line, namely the FTQC threshold line. This match indicates that the quantum computation is fault-tolerant under the condition $C_{\rm int} \geq 1130$. Thus, the FTQC requirements can be estimated from the error map.
%
\section{Finite pulse length}
\label{fin}
\begin{figure}
\includegraphics[clip,width=8cm]{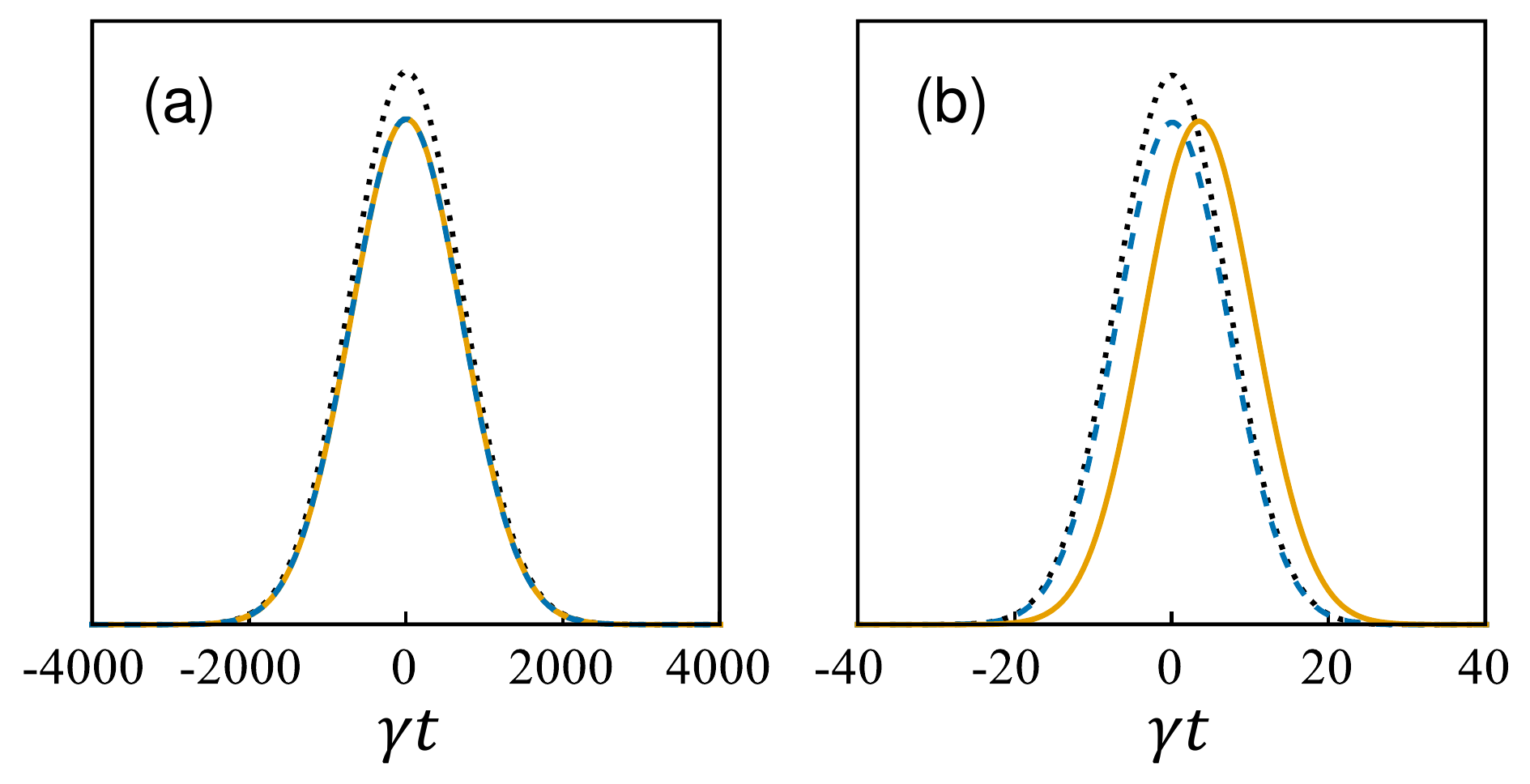}
\caption{Typical shapes of output pulses with the atom in the state $|0\rangle_{\rm a}$ (solid curve) and $|1\rangle_{\rm a}$ (dashed curve). The dotted curves represent input pulses, whose centers are located at $t=0$. The pulse length in (a) is long enough compared with any other characteristic timescales in the system. The pulse length in (a) is two orders of magnitude larger than that in (b). The values of the system parameters other than the pulse length are common to (a) and (b). The distortion of the pulse shape is more noticeable in (b) than in (a).}
\label{dist}
\end{figure}

%
In the case of a finite pulse length, the conditional error increases due to pulse distortion, as shown in Fig.\,\ref{dist}(b), and it changes the situation. In what follows, we discuss the optimization of $\kappa_{\rm ext}$ for pulses with a finite length; then, we compare the quality of cQED required for FTQC for a finite pulse length and in the long pulse limit.

To investigate the dependence of the error on the system parameters, we simulated the dynamics of the CPF gate based on Eqs.\,(\ref{em1})-(\ref{em5}) with a Gaussian input single-photon pulse defined as:
%
\begin{align}
f^{\rm in}(t) = \sqrt{\frac{1}{\sqrt{\pi}W_{\rm t}}}\exp\left(-\frac{(t-t_0)^2}{2W_{\rm t}^2}\right)   .
\end{align}
%
%
%
\subsection{Optimization including the pulse distortion error}
\label{opt2}
For pulses with a finite length, the conditional error caused by distortion of the output-pulse shapes grows up with the unbalanced-photon-loss error. The question here is ``does $\kappa_{\rm ext}^{\rm loss}$ obtained in the long pulse limit also minimize the error due to the pulse distortion?'' As we will see below, the answer is ``no''; $\kappa_{\rm ext}^{\rm loss}$ is not optimal for the total error, i.e., for FTQC. This is because the distortion of the output pulses, which is as serious an issue as the photon loss for short pulses, is a completely different type of error from the photon loss.

Pulse distortion is caused by the reflection depending on the input photon frequency. 
When an input pulse with a finite length is incident, that is, it includes multiple frequency modes besides ${\omega = \omega_{\rm c}}$, a detuning-dependent photon loss and a phase shift between the input and output pulses arise, as indicated from Eqs.\,(\ref{l0}) and (\ref{l1}). The detuning-dependent reflection distorts the output-pulse shapes, as shown in Fig.\,\ref{dist}.

\begin{figure}
\includegraphics[clip,width=8cm]{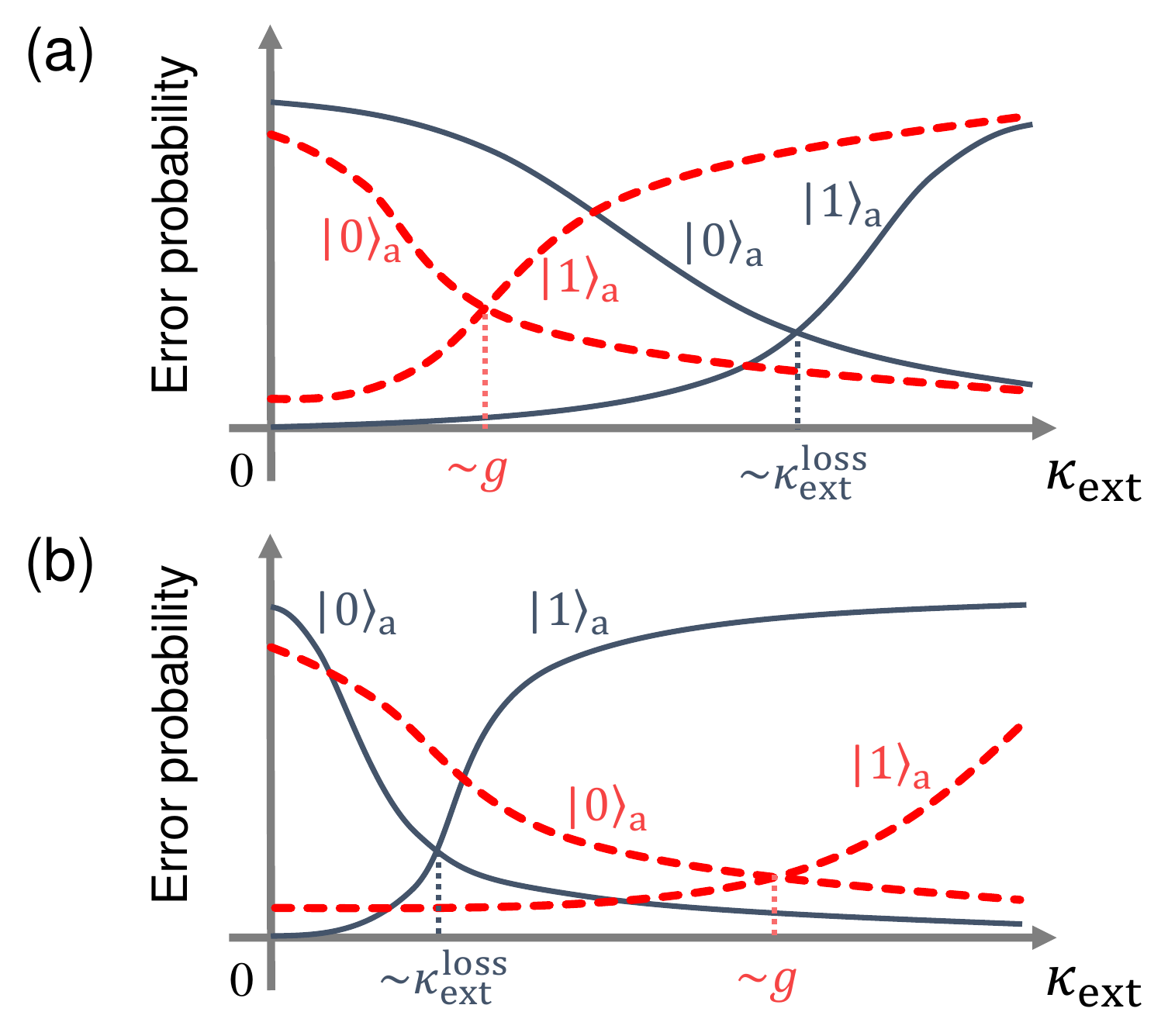}
\caption{Dependence of the error probability due to pulse distortion on $\kappa_{\rm ext}$ (dashed curves) for (a) $\kappa_{\rm int}/\gamma > 1$ and (b) $\kappa_{\rm int}/\gamma < 1$. The solid curves are the conditional error probabilities in the long pulse limit in Fig.\,\ref{err2}(b). The total conditional error probability for a finite pulse length is the sum of the solid and dashed curves.}
\label{deidep}
\end{figure}

To suppress the error probability due to pulse distortion, certain conditions regarding the pulse length must be satisfied in addition to Eq.\,(\ref{hf}).
The condition ${\kappa \gg 1/W_{\rm t}}$ ($W_{\rm t}$ is pulse length) must be satisfied to achieve ${L_0 \simeq -1}$. In addition, to achieve ${L_1 \simeq 1}$, the condition ${g \gg 1/W_{\rm t}}$ must be satisfied in the strong coupling limit (${g \gg \kappa, \gamma}$), while the condition ${g^2/\kappa \gg 1/W_{\rm t}}$ must be satisfied in the bad cavity limit (${\kappa \gg g^2/\kappa \gg \gamma}$). However, the condition ${\kappa \gg 1/W_{\rm t}}$ includes ${g \gg 1/W_{\rm t}}$ in the strong coupling limit, while the condition ${g^2/\kappa \gg 1/W_{\rm t}}$ includes $\kappa \gg 1/W_{\rm t}$ in the bad cavity limit.
In total, to keep the conditional error probability low for pulses with a finite length, we require $\kappa \gg 1/W_{\rm t}$ in the strong coupling limit and ${g^2/\kappa \gg 1/W_{\rm t}}$ in the bad cavity limit. Table\,\ref{tab1} summarizes the conditions under which the photon loss probability and the pulse-distortion error probability are small.
%
%
\begin{table}[H]
\caption{Conditions under which the photon loss probability and the pulse-distortion error probability are small.}
\begin{tabular}{p{10em}p{8em}p{8em}}
\hline \hline
& \hfil photon loss & \hfil pulse distortion \\ \hline 
\hfil
bad cavity limit
& \multirow{2}{*}{$g^2/\gamma \gg \kappa_{\rm ext} \gg \kappa_{\rm int}$} & \hfil $g^2/\kappa \gg 1/W_{\rm t}$ \\
 \hfil
strong coupling limit
& & \hfil $\kappa \gg 1/W_{\rm t}$ \\ \hline \hline
\end{tabular}
\label{tab1}
\end{table}
%
The above discussion leads us to conclude that the optimal value of $\kappa_{\rm ext}$ for short pulses is larger than $\kappa_{\rm ext}^{\rm loss}$ when $\kappa_{\rm int}/\gamma \lesssim 1$, whereas it is smaller than $\kappa_{\rm ext}^{\rm loss}$ when $\kappa_{\rm int}/\gamma \gtrsim 1$.
Figure\,\ref{deidep} shows the dependence of the pulse-distortion error probability (dashed curves), whose illustration is based on the discussion in the previous subsection. These error curves intersect at ${\kappa \sim g}$, where $\kappa$ and $g^2/\kappa$ are equal. Moreover, since ${\kappa_{\rm ext} \gg \kappa_{\rm int}}$ in Eq.\,(\ref{hf}) is also needed for high fidelity, the error curves approximately intersect at ${\kappa_{\rm ext} \sim g}$ (the intersections of the dashed curves in Figs.\,\ref{deidep}(a) and \ref{deidep}(b)).
On the other hand, the photon loss and the unbalanced-photon-loss error probabilities reach a minimum around ${\kappa_{\rm ext} = \kappa_{\rm ext}^{\rm loss} \simeq g\sqrt{\kappa_{\rm int}/\gamma}}$ (the intersections of the solid curves in Figs.\,\ref{deidep}(a) and \ref{deidep}(b)), where ${C_{\rm int} \gg 1}$ is assumed. Hence, the optimal $\kappa_{\rm ext}$ for the photon loss is larger (smaller) than $g$ when $\kappa_{\rm int}/\gamma$ is larger (smaller) than unity. Thus, since the optimal value of $\kappa_{\rm ext}$ for the total error probability is expected to be between $g$ and $\kappa_{\rm ext}^{\rm loss}$, it is larger than in the long pulse limit for ${\kappa_{\rm int}/\gamma \lesssim 1}$ (Fig.\,\ref{deidep}(a)), while it is smaller than in the long pulse limit for ${\kappa_{\rm int}/\gamma \gtrsim 1}$ (Fig.\,\ref{deidep}(b)) \cite{expla2}.
%
%
%
%
\begin{figure}
\includegraphics[clip,width=8cm]{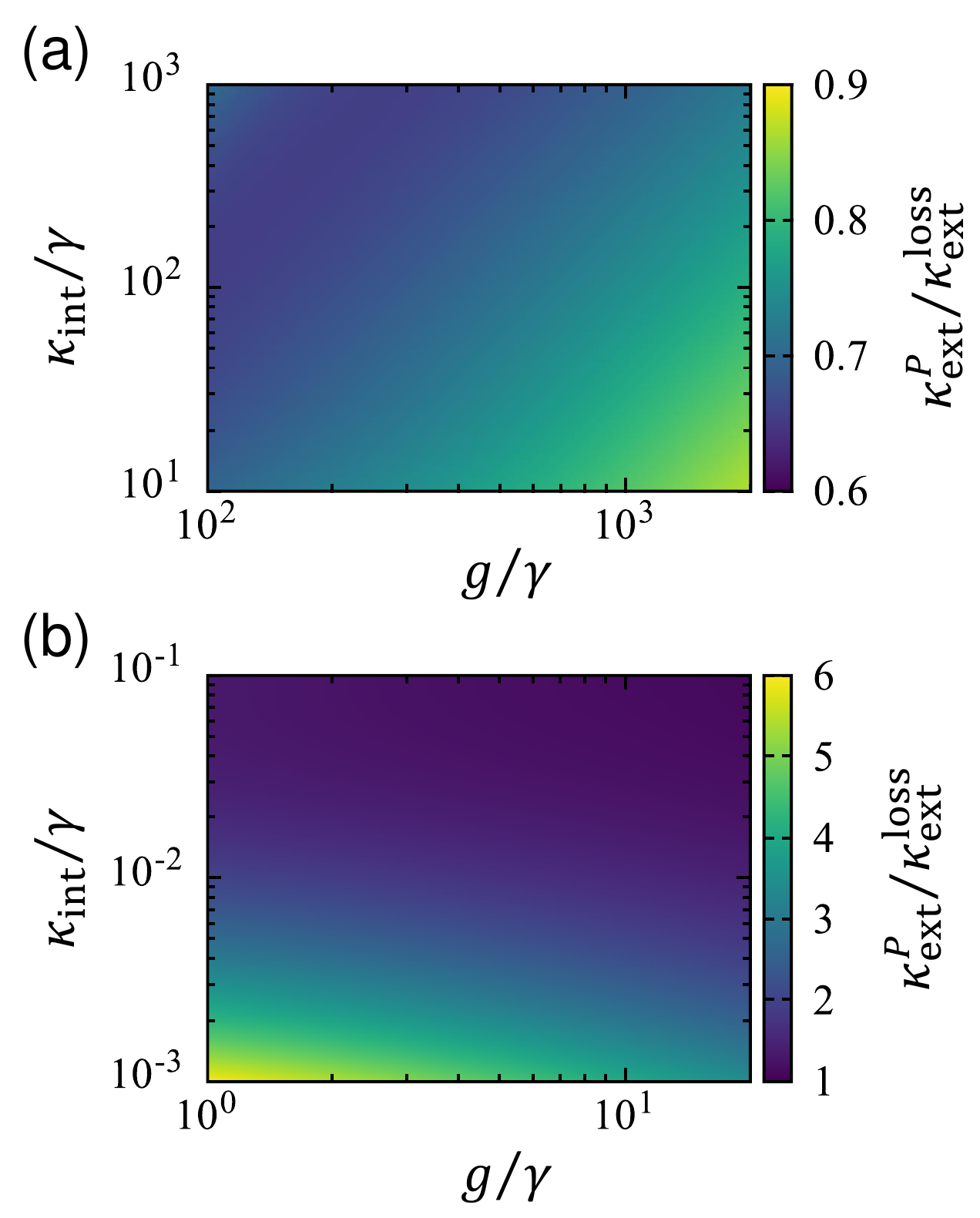}
\caption{Ratio of external coupling rates ${\kappa_{\rm ext}^{P} / \kappa_{\rm ext}^{\rm loss}}$. The pulse length is ${W_{\rm t} =}$ (a) $0.3/\gamma$ and (b) $30/\gamma$.}
\label{comp}
\end{figure}

We performed numerical simulations to confirm the validity of the above discussion. 
In the numerical simulations, the optimal value of $\kappa_{\rm ext}$ for short pulses was determined to minimize the FTQC error parameter $P$. The minimization of $P$ most likely makes the total error enter the fault-tolerant area, which is bounded by the approximate curve $\frac{60}{17}\bar{p_{\rm l}}^{0.59} + \frac{260}{17}\bar{p_{\rm c}}^{0.59} = 1$ of the fault-tolerant threshold. This optimization is the best way for achieving FTQC.

We calculated $\kappa_{\rm ext}^{P}$, which is $\kappa_{\rm ext}$ determined from the minimization of $P$, and plot the ratio ${\kappa_{\rm ext}^{P}/\kappa_{\rm ext}^{\rm loss}}$ in Fig.\,\ref{comp}.
The figure shows that $\kappa_{\rm ext}^{P}$ is small compared with $\kappa_{\rm ext}^{\rm loss}$ for $\kappa_{\rm int}/\gamma > 1$ (Fig.\,\ref{comp}(a)), while it is large for $\kappa_{\rm int}/\gamma < 1$ (Fig.\,\ref{comp}(b)); that is, the optimal $\kappa_{\rm ext}$ for FTQC is larger than in the long pulse limit for ${\kappa_{\rm int}/\gamma \lesssim 1}$, while it is smaller than in the long pulse limit for ${\kappa_{\rm int}/\gamma \gtrsim 1}$. In particular, when $\kappa_{\rm int}/\gamma$ is far from unity, $\kappa_{\rm ext}^{\rm loss}$ is not at all optimal for FTQC with short pulses.
This difference in $\kappa_{\rm ext}$ greatly affects the FTQC requirements, as will be discussed at the end of the next subsection.
%
\subsection{Fault-tolerant quantum computing requirements}
\label{req2}
Here we discuss the FTQC requirements when the pulse length is finite under optimization by minimizing $P$.
We show that the FTQC requirements for large and small $\kappa_{\rm int}/\gamma$ are quite different when the photon pulse is short. This is in contrast to the case of the long pulse limit, where $C_{\rm int} \geq 1130$ is the only FTQC requirement in the whole parameter region. The difference between the FTQC requirements for large and small $\kappa_{\rm int}/\gamma$ comes from the difference between the conditions under which a low pulse-distortion error probability can be achieved for small and large $\kappa_{\rm int}/\gamma$.

\begin{figure}
\includegraphics[clip,width=8cm]{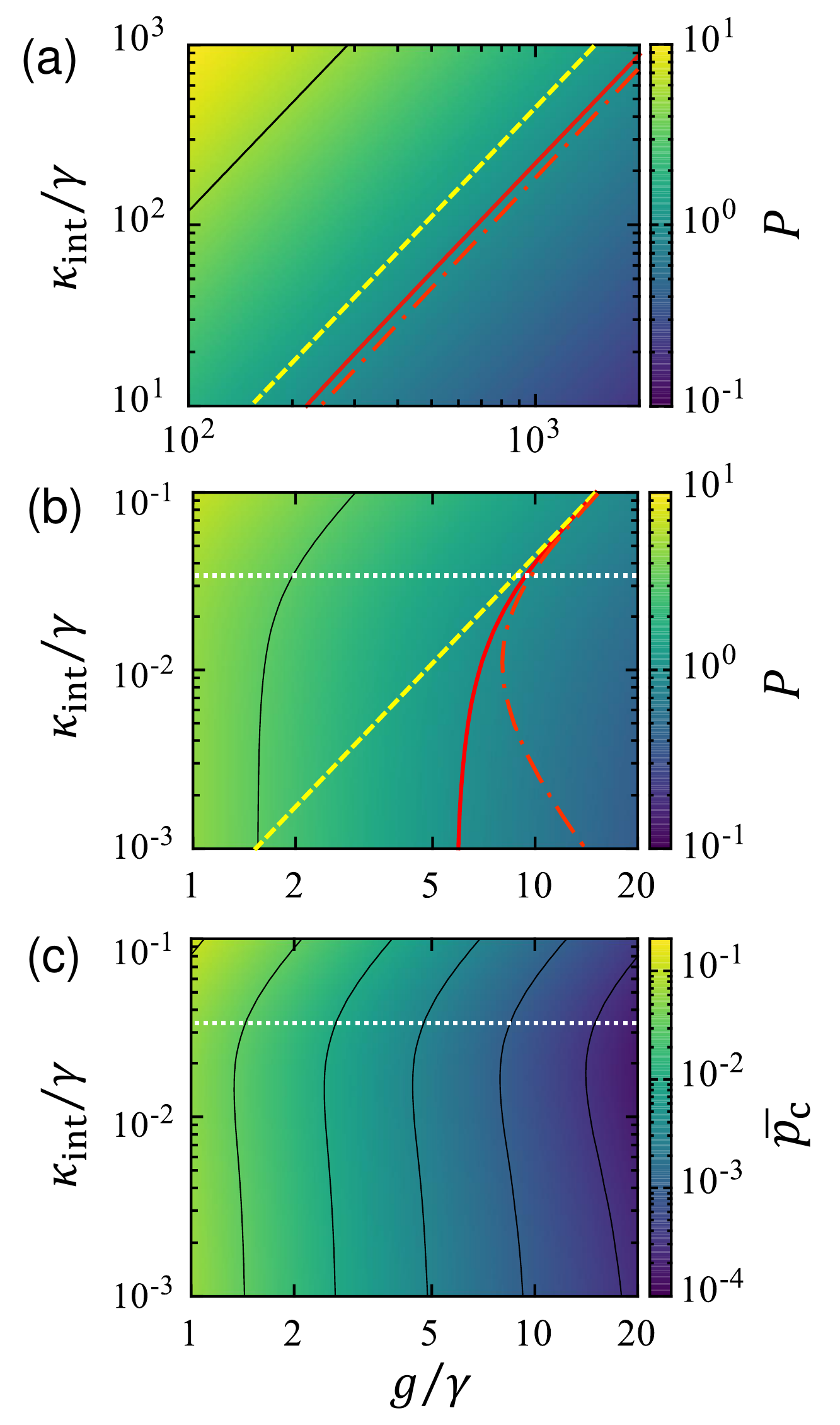}
\caption{(a), (b) FTQC error parameter $P$ and (c) average conditional error probability $\bar{p_{\rm c}}$ in the $(g/\gamma, \kappa_{\rm int}/\gamma)$ plane. The optimization is performed to minimize $P$. The pulse length is (a) ${W_{\rm t} = 0.3/\gamma}$ and (b), (c) ${W_{\rm t} = 30/\gamma}$. The thick solid red line and curve are ${P=1}$ boundaries, namely the FTQC thresholds, and the dashed lines in (a) and (b) are the FTQC thresholds in the long pulse limit as shown in Fig.\ref{gkin}(c). The dashed-dotted curves in (a) and (b) represent the ${P=1}$ boundary when $\kappa_{\rm ext}$ is set to $\kappa_{\rm ext}^{\rm loss}$. The horizontal dotted lines in (b) and (c) denote the ${\kappa_{\rm int}/\gamma = 1/W_{\rm t}}$.}
\label{gkin2}
\end{figure}

Figures\,\ref{gkin2}(a) and \ref{gkin2}(b) plot the FTQC error parameter $P$ in Eq.\,(\ref{ftqcp}) with ${\kappa_{\rm ext} = \kappa_{\rm ext}^{P}}$, when $\kappa_{\rm int}/\gamma$ is large and small. In both parameter regions, the FTQC requirements are more severe than in the long pulse limit because of the pulse distortion error, but the shapes of $P=1$ boundaries are quite different for large and small $\kappa_{\rm int}/\gamma$.

In Fig.\,\ref{gkin2}(a), the $P=1$ line, namely the FTQC threshold line (red line), simply shifts in the direction of increasing $g/\gamma$ (or decreasing $\kappa_{\rm int}/\gamma$) with respect to the FTQC threshold line in the long pulse limit (yellow dashed line). This shift means that the FTQC requirements are still represented by the value of $C_{\rm int}$. The value of $C_{\rm int}$ required for FTQC is $C_{\rm int} \geq 2220$ for $W_{\rm t} = 0.3/\gamma$.
%
%

On the other hand, Fig.\,\ref{gkin2}(b) shows that the $P=1$ boundary is not simply shifted relative to the FTQC threshold line in the long pulse limit: the error probabilities of the CPF gate are not simply determined by $C_{\rm int}$, and consequently the FTQC requirements are not represented by the value of $C_{\rm int}$. The $P=1$ boundary and the FTQC threshold line in the long pulse limit are widely separated for $\kappa_{\rm int}$ smaller than $1/W_{\rm t}$ (the area below the dotted lines) \cite{expla3}. That is, the FTQC requirements are more severe for smaller $\kappa_{\rm int}$. The shape of the ${P=1}$ curve also indicates that reducing $\kappa_{\rm int}$ is less effective to reduce the error parameter $P$ for shorter pulses. It can even cause a situation where the FTQC requirements are never met by reducing $\kappa_{\rm int}$ for small $g$. Thus, unlike in the long pulse limit, increasing $C_{\rm int}$ is not always effective for FTQC, when the photon pulse length is shorter than ${\sim 1/\kappa_{\rm int}}$.

The characteristic shape of the ${P=1}$ boundary in Fig.\,\ref{gkin2}(b) should be considered in developing a strategy to improve a cavity for quantum computation, since typical cavity-QED systems do not have so large $g/\gamma$ (except for artificial atomic systems). The typical values of the cavity-QED parameters are ${(g/\gamma,\kappa_{\rm int}/\gamma)\sim (2.5,0.067)}$ \cite{daiss} and $(5.36,0.04)$ \cite{ruddell,expla4}. In these systems, depending on pulse length, increasing $C_{\rm int}$ by reducing $\kappa_{\rm int}$ is less effective to achieve FTQC. We will show a strategy to alleviate this problem in Sec.\,\ref{rel}.

The difference between the shapes of the ${P=1}$ boundaries in Fig.\,\ref{gkin2}(a) and \ref{gkin2}(b) can be explained by the difference between the conditions to achieve a low pulse-distortion error probability in Fig.\,\ref{gkin2}(a) and \ref{gkin2}(b). Moreover, the difference between the conditions originates from the difference between the parameter regimes of Fig.\,\ref{gkin2}(a) and \ref{gkin2}(b).

In Fig.\,\ref{gkin2}(a), namely for large $\kappa_{\rm int}/\gamma$, the conditions $g \gg \gamma$ and $\kappa_{\rm ext}^{P} > g$ are satisfied because we expect that $g < \kappa_{\rm ext}^{P} < \kappa_{\rm ext}^{\rm loss}$, as explained with the help of Fig.\,\ref{deidep}(a) in the last subsection. Thus, in Fig.\,\ref{gkin2}(a), the cQED parameters satisfy the bad-cavity regime relation, ${\kappa > g^2/\kappa > \gamma}$. In the bad cavity regime, the main condition under which a low pulse distortion error can be achieved is $g^2/\kappa \gg 1/W_{\rm t}$ (see Table\,\ref{tab1}).
On the other hand, in Fig.\,\ref{gkin2}(b), namely for small $\kappa_{\rm int}/\gamma$, $g$ is larger than $\gamma$ and $\kappa_{\rm ext}^{P}$, because it is expected that $\kappa_{\rm ext}^{\rm loss} < \kappa_{\rm ext}^{P} < g$, as explained with the help of Fig.\,\ref{deidep}(b). Thus, in Fig.\,\ref{gkin2}(b), the cQED parameters satisfy the strong-coupling regime relation, ${g \gg \kappa, \gamma}$. In the strong coupling regime, the main condition under which a low pulse distortion error can be achieved is $\kappa \gg 1/W_{\rm t}$ (see Table\,\ref{tab1} again).
Table\,\ref{tab2} summarizes the main conditions under which a low pulse-distortion error probability can be achieved in each parameter region in Figs.\,\ref{deidep}(a) and \ref{deidep}(b).
%
%
\begin{table*}
\caption{Parameter regime and corresponding conditions to suppress pulse-distortion error probability in Fig.\,\ref{gkin2}(a) and \ref{gkin2}(a).}
\begin{tabular}{p{5em}p{12em}p{24em}}
\hline \hline
& \hfil parameter regime & \hfil main condition for low pulse-distortion error probability \\ \hline
\hfil Fig.\,\ref{gkin2}(a) & \hfil bad cavity regime & \hfil $g^2/\kappa \gg 1/W_{\rm t}$ \\
\hfil Fig.\,\ref{gkin2}(b) & \hfil strong coupling regime & \hfil $\kappa \gg 1/W_{\rm t}$ \\ \hline \hline
\end{tabular}
\label{tab2}
\end{table*}

The above discussion simply explains the shift of the FTQC line in the case of the bad cavity regime ( Fig.\,\ref{gkin2}(a)): by satisfying the condition ${g^2/\kappa \gg 1/W_{\rm t}}$, the internal cooperativity $C_{\rm int}$ required for FTQC increases in order to reduce the pulse-distortion error probability, and this results in the shift of the $P=1$ boundary.

The case of the strong coupling regime, i.e., Fig.\,\ref{gkin2}(b), is more complicated. In this regime, as $\kappa_{\rm int}$ decreases, $\kappa_{\rm ext}$ should be increased to satisfy the condition ${\kappa \gg 1/W_{\rm t}}$ for suppressing the pulse distortion error. On the contrary, ${\kappa_{\rm ext}^{\rm loss} = \kappa_{\rm int}\sqrt{1+2C_{\rm int}} \simeq \kappa_{\rm int}\sqrt{2C_{\rm int}} \propto \sqrt{\kappa_{\rm int}}}$ (${C_{\rm int} \gg 1}$ was assumed) decreases as $\kappa_{\rm int}$ is reduced; hence, $\kappa_{\rm ext}$ should be reduced in order to suppress the photon loss probability. Therefore, simply increasing $\kappa_{\rm ext}$ by reducing $\kappa_{\rm int}$ to suppress the pulse distortion error is not the best way of minimizing the total error. Reducing $\kappa_{\rm int}$ ultimately results in an increase in the pulse distortion error when one chooses $\kappa_{\rm ext}$ that minimizes the total error (Fig.\,\ref{gkin2}(c)).
This increase in the pulse distortion error offsets the original decrease in the photon loss by reducing $\kappa_{\rm int}$; thus, reducing $\kappa_{\rm int}$ is less effective for FTQC.
This problem is especially noticeable when $\kappa_{\rm int}$ is smaller than $1/W_{\rm t}$.

Finally, let us address the validity of the optimization minimizing the FTQC error parameter $P$ by comparing it with the optimization only minimizing the photon loss probability, which is valid in the long pulse limit. As can be seen from Fig.\,\ref{gkin2}(b), the optimization minimizing the FTQC error parameter $P$ is especially effective for small $\kappa_{\rm int}/\gamma$, where the pulse distortion error is large. The FTQC with ${\kappa_{\rm ext}=\kappa_{\rm ext}^{\rm loss}}$ requires even more severe conditions to be placed on the cQED parameters compared with our optimization minimizing $P$, for $\kappa_{\rm int}$ smaller than $1/W_{\rm t}$. For example, when ${\kappa_{\rm int}/\gamma = 0.001}$, the value of $g$ required for FTQC with ${\kappa_{\rm ext}=\kappa_{\rm ext}^{\rm loss}}$ is more than twice as large as that with ${\kappa_{\rm ext}=\kappa_{\rm ext}^{P}}$. Thus, appropriately optimizing the external coupling rate according to the FTQC schemes greatly relaxes the FTQC requirements compared with only minimizing the photon loss, especially for small $\kappa_{\rm int}$.
%
\subsection{Optimizing via cavity length}
\label{rel}
We mentioned in the previous subsection that reducing $\kappa_{\rm int}$ is less effective to suppress the errors for shorter pulses for small $\kappa_{\rm int}/\gamma$.
A remaining way to reduce the errors in the parameter region $\kappa_{\rm int} \lesssim 1/W_{\rm t}$ is to increase $g$.
However, we should note that it is not easy to increase only $g$ itself in typical cQED systems \cite{expla5}.
This problem can be alleviated by reducing the cavity length, as following.

In the long pulse limit, as already mentioned, the error probabilities in the CPF gate are characterized only by the internal cooperativity ${C_{\rm int} = g^2/2\kappa_{\rm int}\gamma}$. Assuming that bulk loss of photons (due to absorption and scattering in the cavity medium) is negligible, the system parameters depend on the cavity length $L_{\rm c}$, as $g \propto 1/\sqrt{L_{\rm c}}$, $\kappa_{\rm int} \propto 1/L_{\rm c}$, and ${\gamma = {\rm const}}$ \cite{reiserer2015,goto2019}; that is, $C_{\rm int}$ is independent on $L_{\rm c}$.
Therefore, the cavity length in each system will be determined considering other experimental constraints.
For short pulses, on the other hand, the error probabilities can no longer be characterized by $C_{\rm int}$ alone in the parameter region $\kappa_{\rm int} \lesssim 1/W_{\rm t}$, and they depend on the cavity length. From the dependence of the system parameters on the cavity length, we see that $\bar{p_{\rm l}}$ and $\bar{p_{\rm c}}$ move to the upper right in Fig.\,\ref{gkin2} while $C_{\rm int}$ remains constant, which makes $P$ smaller mainly because of the reduction in the pulse distortion error. Thus, reducing the cavity length is an effective way to suppress the total error for $\kappa_{\rm int} \lesssim 1/W_{\rm t}$.

\begin{figure}
\includegraphics[clip,width=7cm]{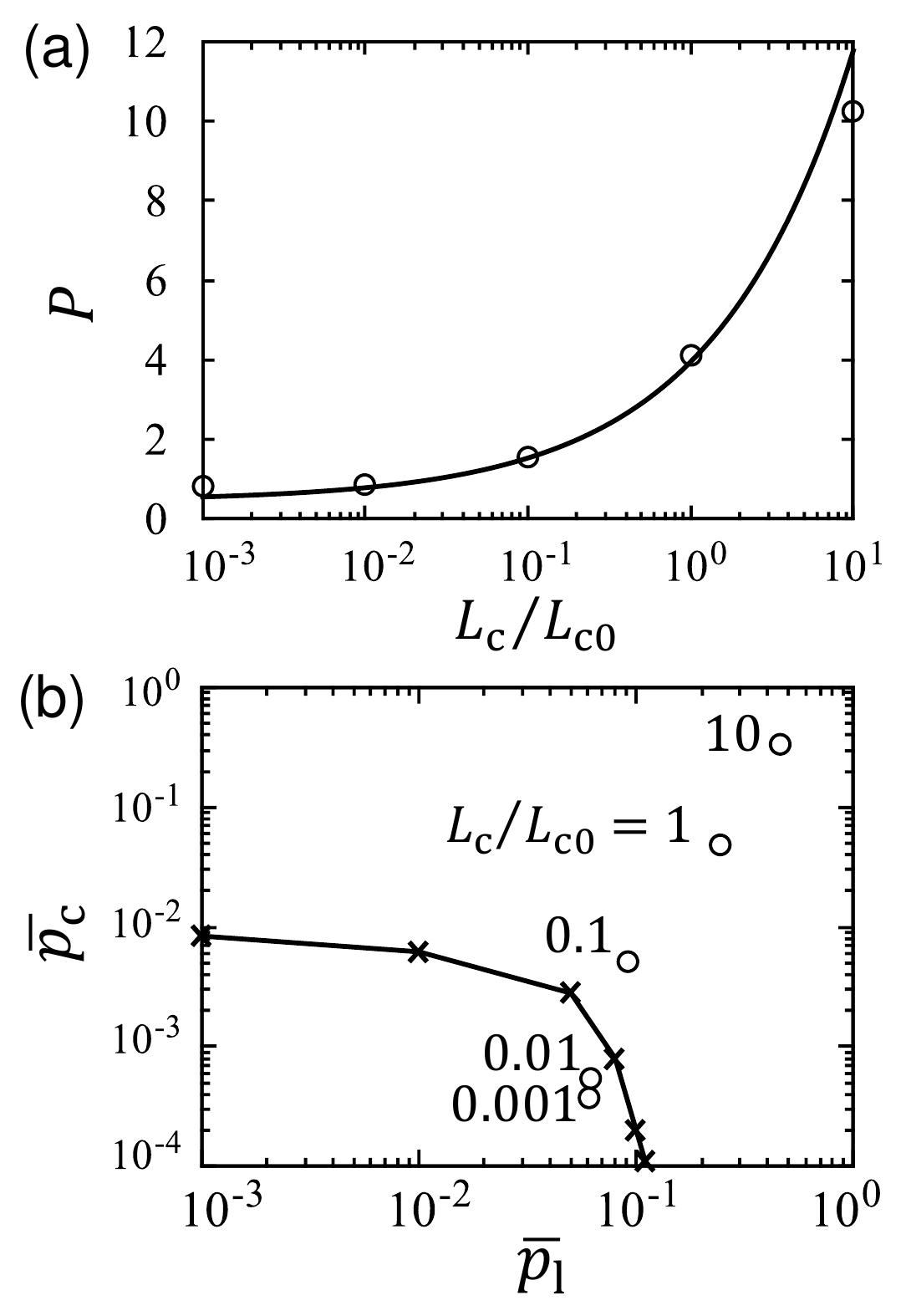}
\caption{(a) Dependence of FTQC error parameter $P$ on cavity length $L_{\rm c}$ for ${W_{\rm t}=30/\gamma}$ (circles). $L_{{\rm c}0}$ represents the cavity length for the system parameter set ${(g/\gamma, \kappa_{\rm int}/\gamma) = (2.0, 0.001)}$. The solid curve represents a fitted function ${P=3.5(L_{\rm c}/L_{{\rm c}0})^{0.51}+0.45}$. (b) Dependence of $\bar{p_{\rm l}}$ and $\bar{p_{\rm c}}$ on the cavity lengths in (a).}
\label{cav}
\end{figure}

To confirm the validity of the above argument, we evaluated the dependence of the average errors on the cavity length $L_{\rm c}$. Figure\,\ref{cav} shows the average errors for ${L_{\rm c}/L_{\rm c0}}$ from $0.001$ to $10$, where $L_{\rm c0}$ denotes the cavity length for the parameters ${(g/\gamma, \kappa_{\rm int}/\gamma) = (2.0, 0.001)}$. We chose these values because the average errors with them fall out of the fault-tolerant area for ${W_{\rm t}=30/\gamma}$, whereas they are in the fault-tolerant area in the long pulse limit (see Fig.\,\ref{gkin2}(d)). In contrast, as shown in Fig.\,\ref{cav}(b), the average errors for ${W_{\rm t}=30/\gamma}$ are in the fault-tolerant area for ${L_{\rm c}/L_{\rm c0} \leq 0.01}$. This indicates that reducing the cavity length can relax the conditions for FTQC with fast gates where $\kappa_{\rm int}$ is smaller than $\sim 1/W_{\rm t}$.
%
%
\section{Summary}
%
In summary, we have described the fault tolerance requirements on an atom-atom CPF gate based on cQED mediated by a photon with a finite pulse length. In our analysis, we optimized the external coupling rate, where two types of the errors caused by photon loss and distortion of the output-pulse shapes were minimized so that the FTQC requirements for the FTQC scheme is most likely to be met. Our optimization greatly relaxes the requirements for FTQC compared with the previous optimization, in which the photon loss probability was simply minimized, especially when the pulse length is shorter than $1/\kappa_{\rm int}$.

In the case of a finite pulse length, the error caused by the distortion of the output-pulse shape increases and the FTQC requirements are more severe, when the timescales $1/\kappa$ and $\kappa/g^2$ are not short enough compared with pulse length in the strong coupling regime and in the bad cavity regime, respectively. Especially for small $\kappa_{\rm int}/\gamma$, reducing the cavity internal loss, i.e., the improving only $\kappa_{\rm int}$, is less effective for FTQC. This is because the optimal external coupling rates to minimize photon loss and the pulse distortion are far apart, resulting in a high total error for any external coupling rate.
However, even in that situation, reducing the cavity length can reduce the total error probability.

In this paper, we focused on the fundamental errors in the gate operation.
A more concrete analysis specifying the atomic system and decoherence effects should be conducted in future.
Our analysis of short photon pulses for obtaining fast gate operations would be helpful in this regard.   


\begin{acknowledgments}
We are grateful to T. Utsugi for useful discussions. This work was supported by JST [Moonshot R\&D][Grant Number JPMJMS2061] and JST CREST, Grant Number JPMJCR1771, Japan.
\end{acknowledgments}


\begin{thebibliography}{99}
\bibitem{expla}
In this study, we focus on the fundamental errors in the gate operation except for decoherence of atom qubits such as atom heating.
\bibitem{expla2}
To ensure the above argument more rigorously, we should discuss whether $\kappa_{\rm ext}^{\rm loss}$ still minimizes the photon loss probability for finite pulse lengths. For finite pulse lengths, the main components of the pulse still have frequencies around ${\Delta = 0}$, and $\kappa_{\rm ext}$ that minimizes the photon loss probability is not far from $\kappa_{\rm ext}^{\rm loss}$. Therefore, we qualitatively discussed the optimal external coupling rate for a finite pulse length using $\kappa_{\rm ext}^{\rm loss}$. Figure\,\ref{comp}, as mentioned below, shows that this qualitative discussion is consistent with the numerical results.
\bibitem{expla3}
Note that we chose the pulse lengths in Figs.\,\ref{gkin2}(a) and \ref{gkin2}(b) so that each deviation between $P=1$ boundary and the FTQC line in the long pulse limit whould be easy to see. For the same pulse length, the $P=1$ boundaries in Figs.\,\ref{gkin2}(a) and \ref{gkin2}(b) would be connected smoothly.
\bibitem{expla4}
These values of the system parameters are estimated supposing that an atom is trapped 100nm away from the nanofiber cavity in Ref.\,\cite{ruddell}. The values change to ${(g/\gamma,\kappa_{\rm int}/\gamma)\sim (3.65,0.044)}$ when an atom is trapped 200nm away from the nanofiber cavity.
\bibitem{expla5}
The atom-field coupling rate $g$ is proportional to $1/\sqrt{A}$, where $A$ is the effective cross-section area of the cavity mode. In typical Fabry-Perot cQED systems, $A$ is already close to the limit.
\end{thebibliography}
\end{document}